\documentclass[aps,pre,showpacs,twocolumn]{revtex4}
\usepackage{bm}
\usepackage{graphicx}
\bibstyle{approve.bib}
\usepackage{amssymb}
\usepackage{amsmath}
\usepackage{esint}
\usepackage{epstopdf}
\usepackage{color}
\usepackage{gensymb}

\newcommand{\be}{\begin{equation}}
\newcommand{\ee}{\end{equation}}
\newcommand{\bea}{\begin{eqnarray}}
\newcommand{\eea}{\end{eqnarray}}
\newcommand{\lan}{\left\langle}
\newcommand{\ran}{\right\rangle}
\newcommand{\br}{\mathbf{r}}

\newcommand{\bk}{\mathbf{k}}

\newcommand{\e}{\varepsilon}

\newcommand{\tv}{\tilde{v}}

\newcommand{\bz}{\bar{z}}
\newcommand{\bd}{\bar{d}}
\newcommand{\pa}{\parallel}

\begin{document}

\title{Electrostatic energy barriers from dielectric membranes upon approach of translocating DNA molecules}

\author{Sahin Buyukdagli$^{1}$\footnote{email:~\texttt{Buyukdagli@fen.bilkent.edu.tr}}  
and T. Ala-Nissila$^{2,3}$\footnote{email:~\texttt{Tapio.Ala-Nissila@aalto.fi}}}
\affiliation{$^{1}$Department of Physics, Bilkent University, Ankara 06800, Turkey\\
$^{2}$Department of Applied Physics and COMP Center of Excellence, Aalto University School of Science, P.O. Box 11000, FI-00076 Aalto, Espoo, Finland\\
$^{3}$Department of Physics, Brown University, Providence, Box 1843, RI 02912-1843, U.S.A.}
\date{\today}

\begin{abstract}
We probe the electrostatic cost associated with the approach phase of DNA translocation events. Within an analytical theory at the Debye-H\"{u}ckel level, we calculate the electrostatic free energy of a rigid DNA molecule interacting with a dielectric membrane. For carbon or silicon based low permittivity neutral membranes, the DNA molecule experiences a repulsive energy barrier between $10$ $k_BT$ and $100$ $k_BT$. In the case of engineered membranes with high dielectric permittivity, the membrane surface attracts the DNA with an energy of the same magnitude. Both the repulsive and attractive interactions result from image-charge effects and their magnitude survive even for the thinnest graphene-based membranes of size $d\sim6$ {\AA}. For weakly charged membranes, the electrostatic free energy is always attractive at large separation distances but switches to repulsive close to the membrane surface. We also characterise the polymer length dependence of the interaction energy. For specific values of the membrane charge density, low permittivity membranes repel short polymers but attract long polymers. Our results can be
used to control the strong electrostatic free energy of DNA-membrane interactions prior to translocation events by chemical
engineering of the relevant system parameters.
\end{abstract}

\date{\today}
\maketitle

\section{Introduction}

DNA is the most important mediator of biological information during assembly of 
the building blocks of living organisms. As the carrier of the genetic code, DNA plays a central role in various biological and technological processes such as cell division~\cite{b1}, protein biosynthesis~\cite{b2}, drug delivery~\cite{b3}, and DNA profiling~\cite{b4}. The efficient use of DNA in biological and nanotechnological applications necessitates a fast access to its genetic content and an accurate knowledge of its interaction with the surrounding medium. Considering the omnipresent coupling between strongly charged DNA molecules, the dielectric water solvent embodying charges, and external macromolecules/membranes in Nature, 
a proper modelling of DNA electrostatics becomes essential. 

A fundamental question concerning DNA in biological and artificial systems concerns the electrostatic interactions between fluctuating polymers and membranes. This has been mainly considered at the mean-field (MF) Poisson-Boltzmann (PB) level. The electrostatic MF approximation has the advantage of allowing the consideration of entropic polymer fluctuations. The corresponding formalism consists of coupling Edward's path integral model~\cite{ed} with the field theoretic Coulomb liquid model~\cite{pod0}. In this direction, one can mention the seminal works of Podgornik ~\cite{pod1,pod2}, where he considered the electrostatics of an infinitely long polyelectrolyte  between two charged membrane walls. Within the same MF approximation, the interaction of a polyelectrolyte with a charged sphere was considered in Ref.~\cite{dun} and possible extensions beyond MF level were proposed. Similar MF approaches have been subsequently applied to polyelectrolyte brushes~\cite{orland} and  polymer-interface interactions in incompressible liquids~\cite{muthu}. 

Electrohydrodynamic theories of confined ions and polymers beyond the MF approximation have been developed for rigid polyelectrolytes. In Ref.~\cite{Buyuk2014}, we coupled one-loop electrostatic equations with the Stokes equation and calculated the electrophoretic DNA mobility and ionic currents in confined pores. Within this theory that accounts for charge correlations associated with the low membrane permittivity and charge multivalency, we showed that the addition of multivalent counterions into the solution reverses the MF electrophoretic mobility of polyelectrolytes. It is noteworthy that this effect was recently observed in electrophoretic DNA transport experiments~\cite{soft}. Then, by applying the theory to hydrodynamically induced DNA transport, we found that during polymer translocation events, the multivalency induced charge correlations reverse the ionic current through neutral pores~\cite{Buyuk2015}. 

An important feature of the correlation-corrected polymer transport theories is that they neglect the interaction between the membrane and the portion of the DNA located outside the nanopore. In the present article, we address this issue by considering the electrostatic free energy of a polyelectrolyte located outside a dielectric membrane. Our theory aims at quantitatively evaluating the electrostatic cost, i.e. the electrostatic contribution to the free energy barrier, upon the approach phase preceding DNA translocation events. Understanding how to control this barrier is paramount to successful applications of DNA translocation.

At this point, we should also mention the important beyond-MF models of Refs.~\cite{corrpol1,corrpol2} where the effect of polarization charges on polymer adsorption onto planar interfaces was considered. The major approximation of these theories consists of replacing the electrostatic many-body potential by a one-body image-charge potential in the path integral over polymer configurations. In order to avoid the resulting uncontrollable errors and to simplify the theoretical framework, we consider here a rigid polyelectrolyte approaching a charged dielectric membrane. In the beginning of Section~\ref{sec}, we calculate the electrostatic free energy of the polymer induced by the presence of the membrane. Section~\ref{neut} is devoted to neutral membranes. We scrutinize the effect of the polymer length, salt density, and membrane thickness and permittivity on the polyelectrolyte free energy. Then, in Section~\ref{charged}, we consider a charged membrane and investigate the competition between image charge and membrane surface charge forces in polymer-membrane interactions. The limitations and possible extensions of our theory are discussed in the Summary and Conclusions part.

\section{Debye-H\"uckel theory of polymer-membrane interactions}
\label{sec}
\begin{figure}
\includegraphics[width=1\linewidth]{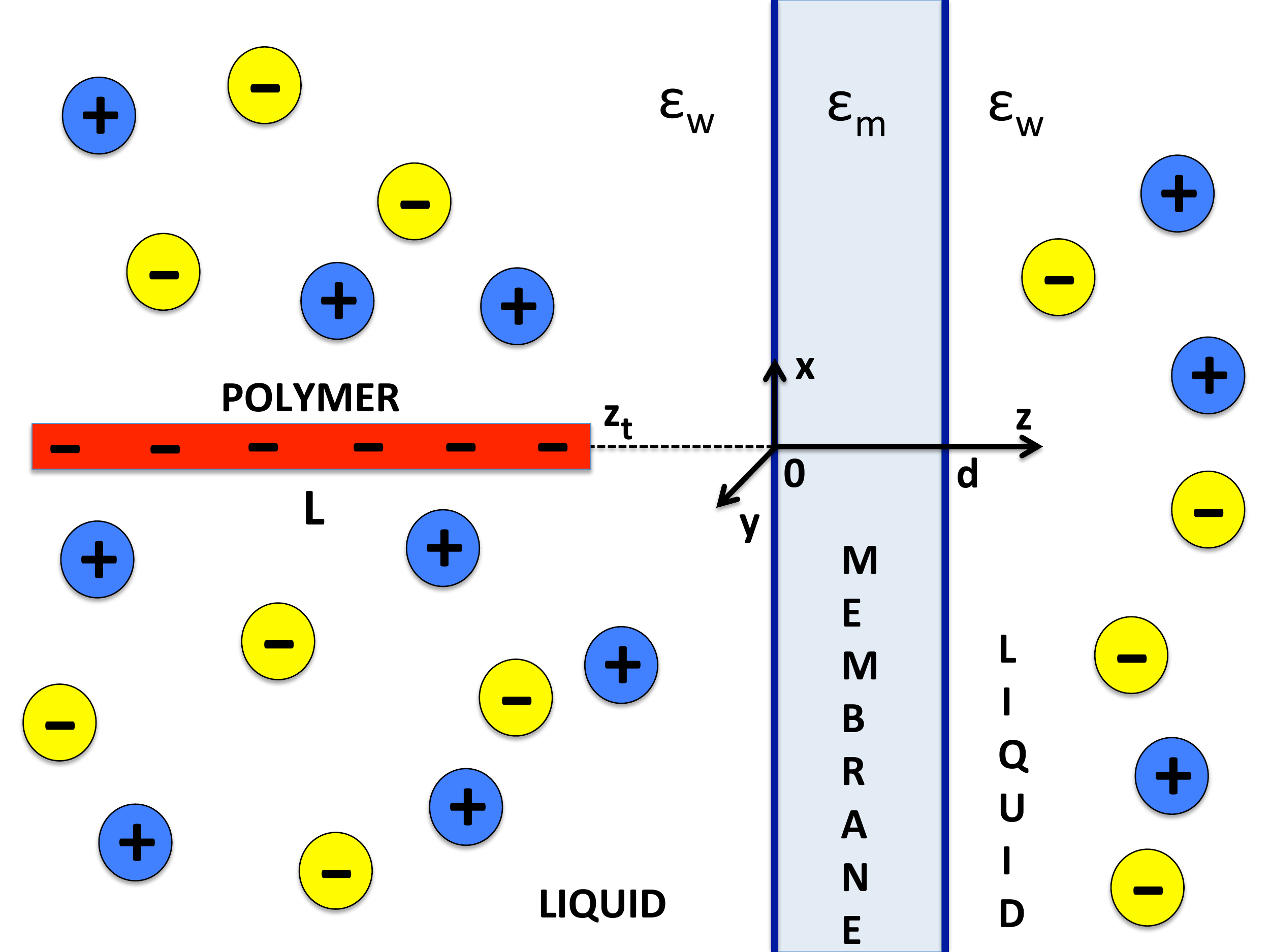}
\caption{(Color online) Polyelectrolyte of length $L$ and linear charge density $-\lambda<0$ whose right end is located at a distance of $z=z_t<0$ from the membrane. The membrane has thickness $d$ and dielectric permittivity $\e_m$. Both the polymer and the membrane are immersed in a symmetric monovalent electrolyte solution with bulk concentration $\rho_b$ and dielectric permittivity $\e_w=80$.}
\label{fig1}
\end{figure}

First, we introduce the theoretical model of electrostatic interactions between a DNA molecule and a dielectric membrane 
modelled as in Fig.~\ref{fig1}. The membrane is assumed to consists of two infinite lateral surfaces on the $x-y$ plane, separated by $d$ which is the membrane thickness. The left ($z<0$) and the right lateral surfaces ($z>d$) are in contact with a salt solution. The polyelectrolyte modelled as a rigid line charge of length $L$ is located on the left side of the membrane. In Appendix~\ref{ap1}, we show that the electrostatic Debye-H\"{u}ckel (DH) free energy of the polyelectrolyte is
\be\label{fr1}
\Omega_{\rm pol}=k_BT\int\frac{\mathrm{d}\br\mathrm{d}\br'}{2}\sigma(\br)v_{\rm DH}(\br,\br')\sigma(\br'), 
\ee
where $\sigma(\br)$ is the distribution of the fixed charges (other than the mobile ions), and the potential $v_{\rm DH}(\br,\br')$ is the solution of the DH Eq.~(\ref{ker}) introduced in Appendix~\ref{ap1}. 

For the line charge perpendicular to the membrane, the total charge distribution can be expressed in the form
\be\label{ch1}
\sigma(\br)=-\lambda\;\delta\left(\br_\pa\right)\;g(z)+\sigma_s\delta(z),
\ee
where $\lambda>0$ is the linear DNA charge density, $\br_\pa$ is the vector indicating the position of any point in the $x-y$ plane that coincides with the lateral membrane surface, and $g(z)$ stands for the polymer structure factor along the $z$ axis. In the present work, we will assume that the membrane surface charge of uniform density $-\sigma_s<0$ is located at $z=0$ and the second surface at $z=d$ is neutral. Furthermore, due to the translational symmetry in the membrane plane, one can Fourier expand the Green's function as
\be\label{pot1}
v_{\rm DH}(\br,\br')=\int\frac{\mathrm{d}^2\bk}{4\pi^2}\;e^{i\bk\cdot\left(\br_\pa-\br_\pa'\right)}\tv_{\rm DH}(z,z').
\ee
By inserting into the right-hand-side of Eq.~(\ref{fr1}) the function~(\ref{ch1}) together with the Fourier expansion~(\ref{pot1}) and evaluating the integrals over the membrane surface, the free energy takes the form
\bea
\label{fr2}
\frac{\Omega_{\rm pol}}{k_BT}&=&\lambda^2\int_0^\infty\frac{\mathrm{d}kk}{4\pi}\iint_{-\infty}^{+\infty}\mathrm{d}z\mathrm{d}z'g(z)\tv_{\rm DH}(z,z')g(z')\nonumber\\
&&-\lambda\sigma_s\int_{-\infty}^{\infty}\mathrm{d}zg(z)\tv_{\rm DH}(z,z'=0;k=0).
\eea
In Eq.~(\ref{fr2}), we omitted the membrane self-energy $\Omega_{\rm mem}=\int_{\br,\br'}\sigma_s(\br)v_{\rm DH}(\br,\br')\sigma_s(\br')/2$. In the rest of the article, we will consider a symmetric electrolyte composed of two monovalent species on each side of the membrane, with valencies $q_+=-q_-=1$ and bulk densities $\rho_{+b}=\rho_{-b}=\rho_b$. Moreover, the liquid temperature will be set to the ambient temperature of $T=300$ K, and dielectric permittivities will be expressed in units of the vacuum permittivity $\e_0$.

\subsection{Neutral membranes}
\label{neut}

Next, we will consider the interaction between the polyelectrolyte and a neutral membrane ($\sigma_s=0$). To this aim, we will calculate the net energetic cost for the polyelectrolyte to approach the membrane. In the configuration of the polymer of length $L$ whose right end is located at the distance $z_t\leq0$ from the membrane (see Fig.~\ref{fig1}), the structure factor is given by 
\be\label{str}
g(z)=\theta(z_t-z)\theta(z-z_t+L),
\ee
where $\theta(x)$ is the Heaviside step function. We insert this structure factor into Eq.~(\ref{fr2}) together with the Fourier transformed Green's functions~(\ref{ker3})-(\ref{ker5}) given in Appendix~\ref{ap2}, and subtract the electrostatic bulk free energy associated with the bulk potential $\tv_b(z-z')$ of Eq.~(\ref{ker6}). After carrying out the spatial integrals and noting that the second term of Eq.~(\ref{fr2}) vanishes for $\sigma_s=0$, we get the net electrostatic free energy mediated exclusively by the dielectric membrane in the form
\bea
\label{cost1}
\frac{\Delta\Omega_{\rm pol}(z_t)}{k_BT}&=&\frac{\ell_B\lambda^2}{2}\int_0^\infty\frac{\mathrm{d}kk}{p^3}\frac{\Delta\left(1-e^{-2kd}\right)}{1-\Delta^2e^{-2kd}}\\
&&\hspace{2.0cm}\times\left(1-e^{-pL}\right)^2e^{-2p \vert z_t \vert}.\nonumber
\eea
The electrostatic free energy of Eq. (\ref{cost1}) corresponds to the work done adiabatically to drive the polymer from the bulk region at $z=-\infty$ to the distance $z_t$ from the membrane surface. In Eq.~(\ref{cost1}), we introduced the Bjerrum length $\ell_B=e^2/(4\pi\e_wk_BT)\approx 7$ {\AA} with $\e_w= 80$ being the solvent permittivity, the auxiliary function $p=\sqrt{k^2+\kappa^2}$, where $\kappa^2=8\pi q^2\ell_B\rho_b$ stands for the DH screening parameter, and the dielectric discontinuity function $\Delta=(\e_wp-\e_mk)/(\e_wp+\e_mk)$. Moreover, the delta symbol on the l.h.s. of Eq.~(\ref{cost1}) means that we neglected the bulk contribution and took into account exclusively the energy due to the presence of the membrane. Indeed, we note that in the limit of a bulk electrolyte, i.e. as the membrane thickness tends to zero $d\to0$, the net free energy vanishes, that is $\Delta\Omega_{\rm pol}(z_t)\to0$ due to the membrane's neutrality assumption. 
\begin{figure}
\includegraphics[width=1.2\linewidth]{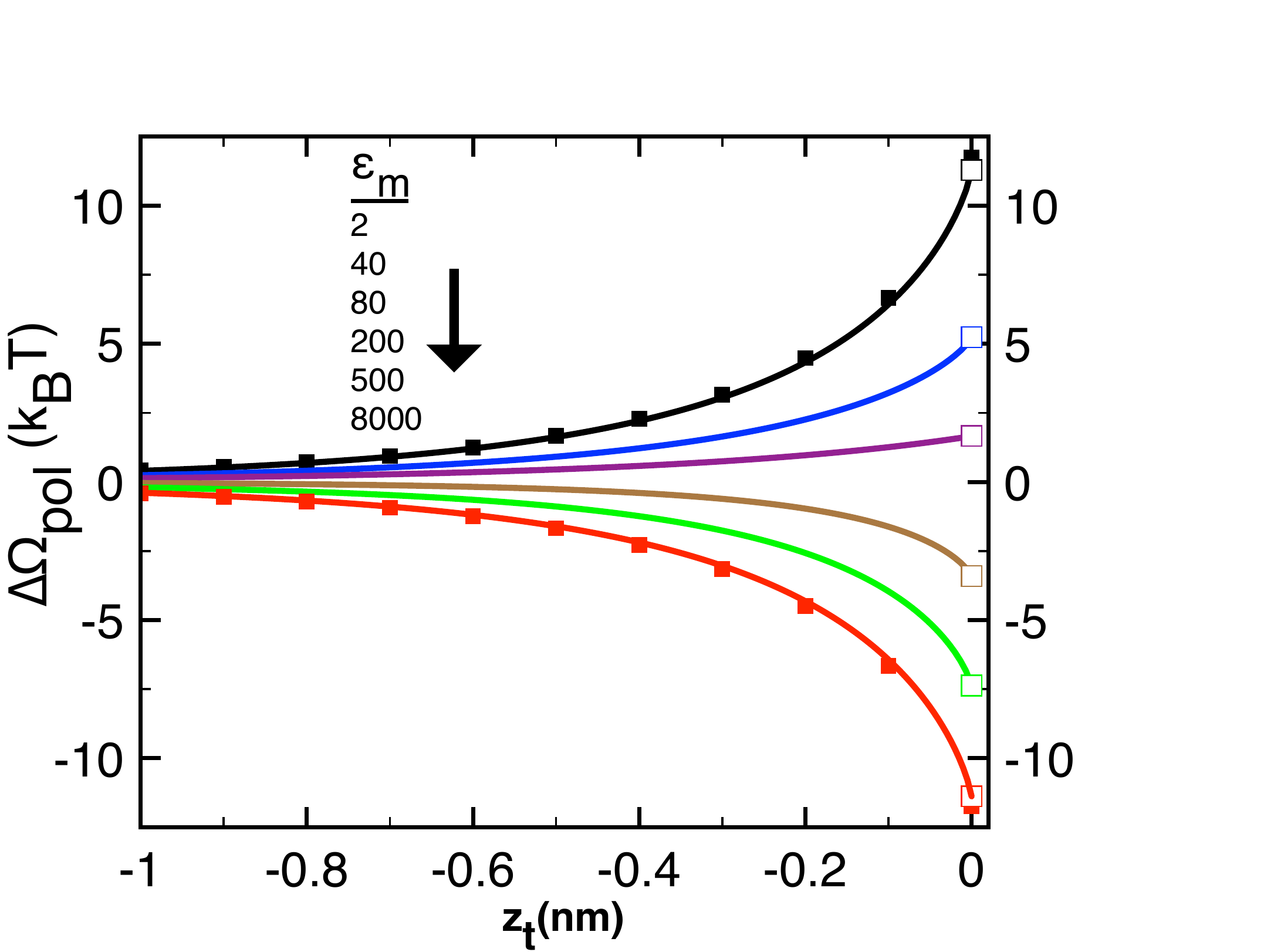}
\caption{(Color online) Electrostatic free energy of Eq.~(\ref{cost1}) against the polymer distance for various membrane permittivities displayed in the legend (solid curves). Bulk ion density is $\rho_b=0.1$ M, pore size $d=10$ nm, polymer length $L=1$ $\mu$m, and DNA charge density $\lambda=2\;e/(0.34\;\mbox{nm})$. Open symbols denoting the energy barrier at $z_t=0$ are from Eq.~(\ref{costIII0}). Black and red symbols correspond to the closed-form expression of Eq.~(\ref{cost2}) with $\e_m=0$ ($s=+1$) and $\e_m=\infty$ ($s=-1$), respectively.}
\label{fig2}
\end{figure}

\subsubsection{Membrane permittivity $\e_m$}
\label{vsem}

The biological and synthetic membranes used in DNA translocation experiments are usually made of carbon or silicon. Such membranes are characterized by a low dielectric permittivity $\e_m\approx 2-8$. However, recent membrane engineering techniques based on the insertion of carbon structures or graphene nanoribbons (GNRs) into Si-based host matrices allow to increase the permittivity of these materials up to $8000$~\cite{Dimiev,Dang}. In order to predict electrostatic membrane-polymer interactions over the experimentally relevant permittivity range, we plot in Fig.~\ref{fig2} the electrostatic free energy of Eq.~(\ref{cost1}) for a polymer of length $L=1$ $\mu$m against its distance $z_t$ from the membrane for various permittivity values. The charge density is set to the linear charge density of dsDNA, that is $\lambda=2\;e/(0.34\;\mbox{nm})$. The other model parameters are given in the figure caption.

In Fig.~\ref{fig2}, for C/Si-based membranes with small permittivities ($\e_m=2$), the electrostatic free energy of the approaching polymer increases from zero to about $11$ $k_BT$ within about $1$ nm distance.  The reduction of the barrier with increasing membrane permittivity from top to bottom shows that this energetic cost is mainly due to the interaction of the polymer charges with their electrostatic images. For the permittivity value $\e_m=\e_w=80$, where the dielectric discontinuity between the liquid and the membrane vanishes, the barrier survives but its value is reduced by an order of magnitude to $\Delta\Omega_{\rm pol}(0)\approx 2.0\;k_BT$. In the latter case where image-charge interactions are absent, the small barrier is solely due to the electrostatic screening deficiency of the charge-free membrane. Moreover, for GNRs type membranes with a large permittivity $\e_m>\e_w$, the electrostatic free energy become {\it negative}. In other words, similar to point charges at metallic interfaces~\cite{Buyuk2010}, as the membrane dielectric permittivity exceeds that of water, the polymer-membrane interaction switches from repulsive to attractive. For the highest permittivity value $\e_m=8000$ measured for GNRs~\cite{Dang}, the depth of the attractive well reaches a remarkably large value of
$\Delta\Omega_{\rm pol}(0)\approx -11.0\;k_BT$.

We focus next on the electrostatic free energy  at $z_t=0$.  In order to derive an analytical expression, we consider the limit where  the polymer length and the pore thickness tend to infinity, i.e. $L\to\infty$ and $d\to\infty$. 
The physical conditions that validate these limits will be determined below. In these limits, Eq.~(\ref{cost1}) simplifies as
\be\label{costII}
\lim_{L,d\to\infty}\frac{\Delta\Omega_{\rm pol}(0)}{k_BT}=\frac{\ell_B\lambda^2}{2}\int_0^\infty\frac{\mathrm{d}kk}{p^3}\Delta.
\ee
Carrying out the integral and introducing the dielectric contrast parameter $\gamma=\e_m/\e_w$, the free energy takes the form
\be\label{costIII0}
\lim_{L,d\to\infty}\frac{\Delta\Omega_{\rm pol}(0)}{k_BT}=\frac{\ell_B\lambda^2}{2\kappa}F(\gamma),
\ee
with the auxiliary function
\bea
\label{costIII}
F(\gamma)&=&-1+\frac{\pi}{\gamma}-\frac{2}{\gamma}\frac{\arccos(\gamma)}{\sqrt{1-\gamma^2}}, \hspace{3mm}\mathrm{for}\hspace{3mm}\gamma<1;\\
\label{costIV}
F(\gamma)&=&\pi-3,\hspace{3mm} \mathrm{for}\hspace{3mm}\gamma=1;\\
\label{costV}
F(\gamma)&=&-1+\frac{\pi}{\gamma}-\frac{2}{\gamma}\frac{\ln\left[\gamma+\sqrt{\gamma^2-1}\right]}{\sqrt{\gamma^2-1}},\hspace{3mm}\mathrm{for}\hspace{3mm}\gamma>1.\nonumber\\
\eea
In Fig.~\ref{fig2}, we show that the simple law of Eq.~(\ref{costIII0}) accurately reproduces the electrostatic free energy at $z_t=0$ for various membrane permittivities (open square symbols at zero distance). 

\begin{figure}
\includegraphics[width=1.2\linewidth]{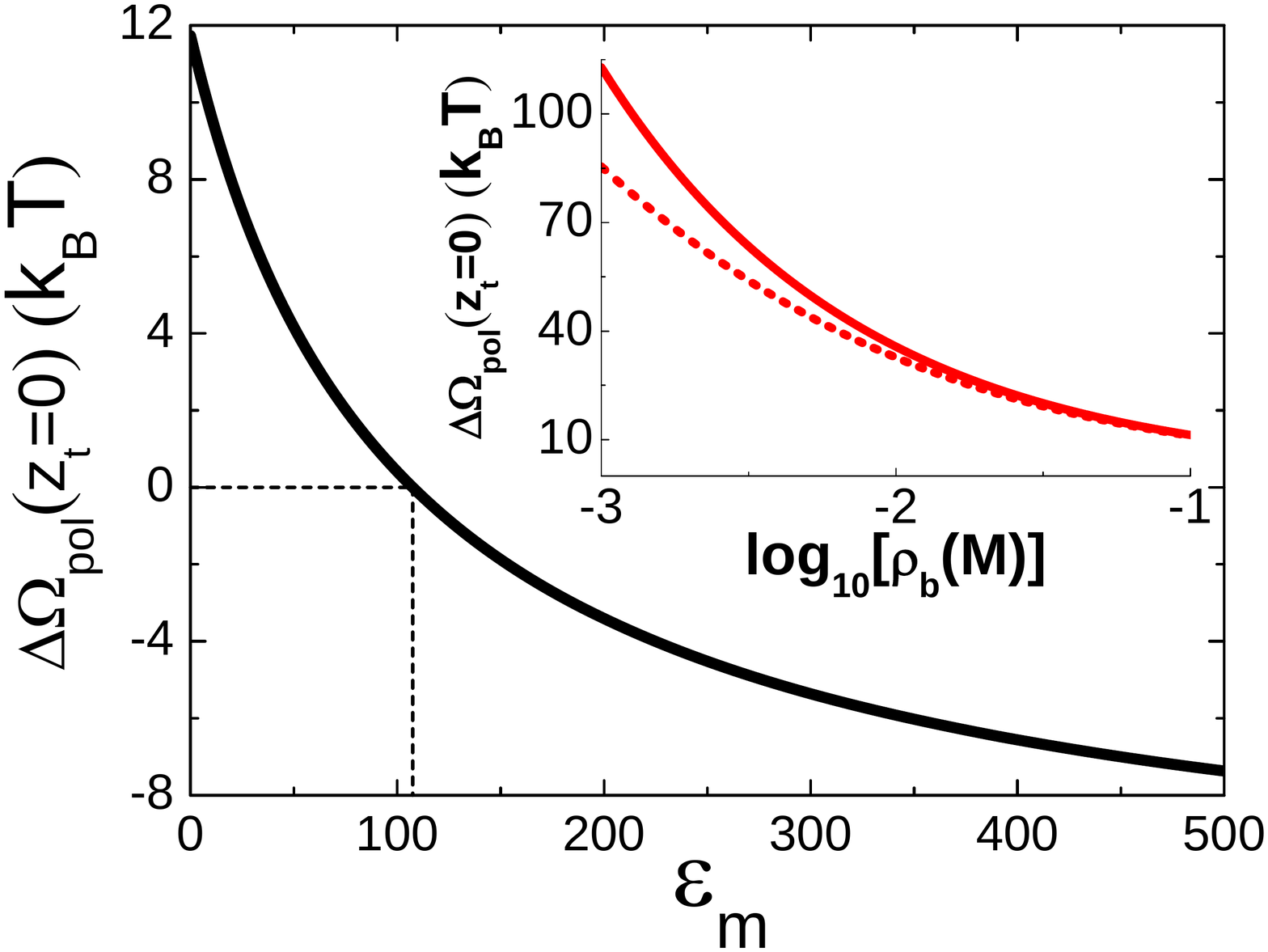}
\caption{(Color online) Electrostatic free energy of Eq.~(\ref{costIII0}) at the membrane surface against membrane permittivity $\e_m$ at density $\rho_b=0.1$ M (main plot) and salt concentration $\rho_b$ at permittivity $\e_m=2$ (solid curve in the inset). The dashed curve in the inset obtained from Eq.~(\ref{cost1}) for $L\to\infty$, $\e_m=2$, and $d=6$ {\AA} generalizes the result in the solid curve to a finite membrane thickness. The remaining model parameters are the same as in Fig.~\ref{fig2}.}
\label{fig2II}
\end{figure}

In the main plot of Fig.~\ref{fig2II}, we plot the electrostatic free energy of Eq.~(\ref{costIII0}) at the membrane surface versus the membrane permittivity. In agreement with Fig.~\ref{fig2}, with an increase of the permittivity from $\e_m=2$ to 500, the free energy is seen to evolve from $+12$ $k_BT$ to $-8$ $k_BT$. As indicated by the dashed lines in the same figure, the electrostatic free energy switches from repulsive to attractive at the permittivity value $\e_m \approx 107$, where the weak attractive image force exactly compensates for the repulsive solvation force induced by the charge screening deficiency of the membrane. In the next subsection we scrutinize the polymer length and salt dependence of this interaction energy.

\subsubsection{Polymer length $L$ and salt density $\rho_b$}
\label{LvsRHO}

DNA translocation experiments are carried out with different sequence lengths and salt concentrations. Motivated by this point, we focus now on the salt and polymer length dependence of the DNA-membrane interactions. To this aim, we will derive a closed-form expression for the electrostatic free energy profile of Eq.~(\ref{cost1}) in the case of very low and very large permittivity membranes.  First, we introduce an auxiliary parameter $s$ that will allow to cover the case of biological or silicon-based membranes of low permittivities ($\e_m\ll\e_w$) and engineered membranes including GNRs of large permittivities ($\e_m\gg\e_w$)~\cite{Dimiev},
\bea\label{par}
\label{con1}
s&=&+1,\hspace{3mm}\mathrm{for}\hspace{3mm}\e_m=0\hspace{2mm}\mathrm{(bio/Si\hspace{1mm}membranes);}\\
\label{con2}
s&=&-1,\hspace{3mm}\mathrm{for}\hspace{3mm}\e_m=\infty\hspace{2mm}\mathrm{(GNRs).}
\eea
In the upper and lower limits defined by Eqs.~(\ref{con1})-(\ref{con2}), the dielectric discontinuity function $\Delta$ in Eq.~(\ref{cost1}) tends to $+1$ and $-1$, respectively, which allows to carry out the Fourier integral. We find
\be\label{cost2}
\Delta\Omega_{\rm pol}(z_t)=sk_BT\frac{\ell_B\lambda^2}{2\kappa}G(z_t),
\ee
where we defined the adimensional function
\bea\label{g}
G(z_t)&=&e^{2\kappa z_t}+e^{-2\kappa(L-z_t)}-2e^{-\kappa(L-2z_t)}\\
&&-2\kappa z_t\;\mathrm{Ei}[2\kappa z_t]+2\kappa(L-z_t)\mathrm{Ei}\left[-2\kappa(L-z_t)\right]\nonumber\\
&&-2\kappa(L-2z_t)\mathrm{Ei}\left[-\kappa(L-2z_t)\right].\nonumber
\eea
In Eq.~(\ref{g}), the exponential integral function is denoted by $\mathrm{Ei}(x)$~\cite{math}. We display the potential of Eq.~(\ref{cost2}) in Fig.~\ref{fig2} by solid square symbols. We note that this analytical form accurately reproduces the energy profile for low permittivity ($\e_m=2$) and large permittivity ($\e_m=8000$) membranes. Using the closed-form expression of Eq.~(\ref{cost2}), we will next scrutinize the dependence of the electrostatic free energy on the polymer length and ion concentration.

\begin{figure}
\includegraphics[width=1.05\linewidth]{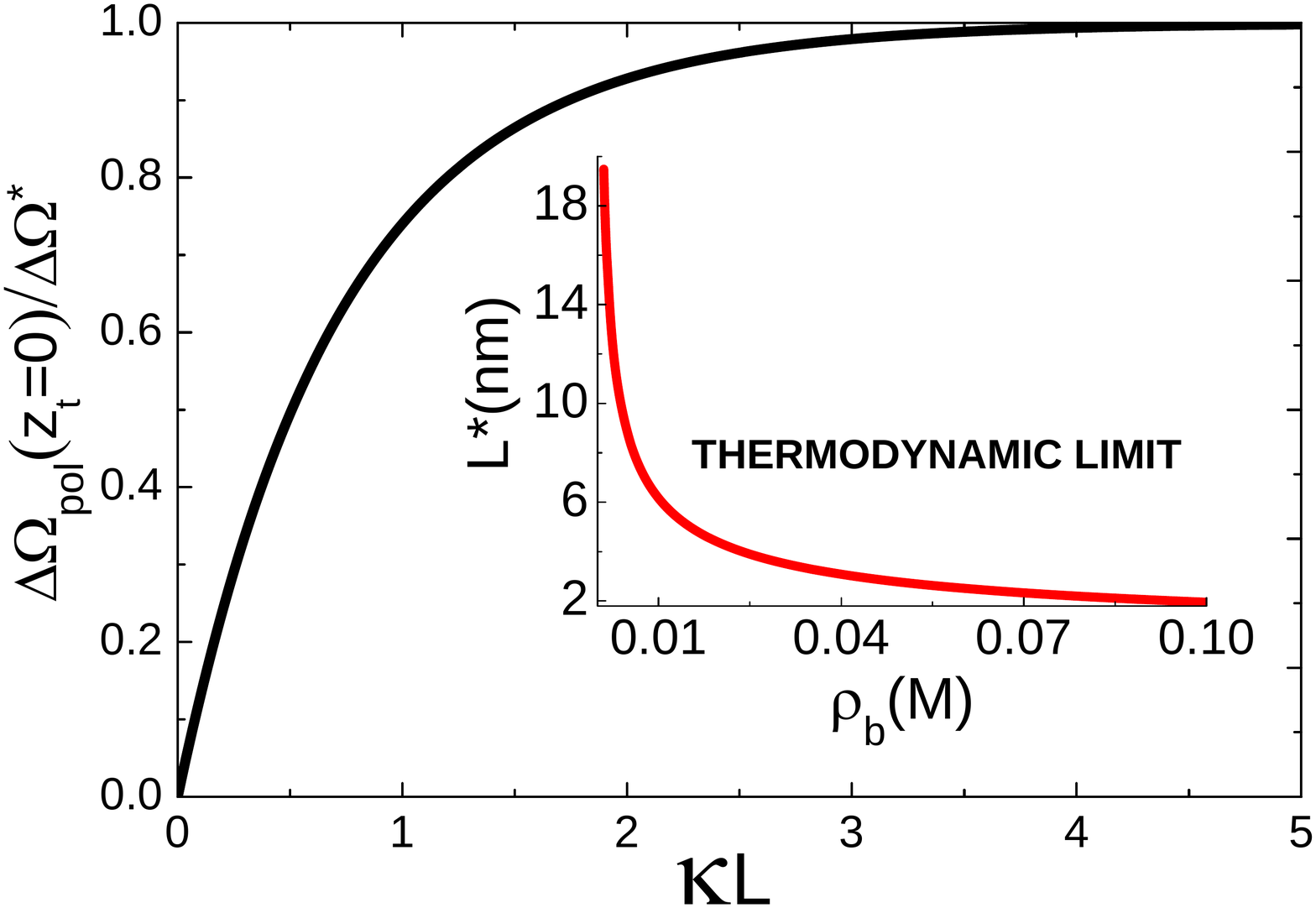}
\caption{(Color online) Main Plot: Rescaled electrostatic free energy of Eq.~(\ref{cost3}) at the membrane surface against the reduced polymer length $\kappa L$ at the membrane permittivity $\e_m=0.0$. Inset: The characteristic polymer length $L^*=2/\kappa$ above which the thermodynamic limit is reached (area above the curve) against the bulk salt density.}
\label{fig3}
\end{figure}

In Fig.~\ref{fig3}, we display the polymer length dependence of the electrostatic free energy Eq.~(\ref{cost2}) at the membrane surface
\be\label{cost3}
\frac{\Delta\Omega_{\rm pol}(0)}{s\Delta\Omega^*}=\left(1-e^{-\kappa L}\right)^2+2\kappa L\left[\mathrm{Ei}(-2\kappa L)-\mathrm{Ei}(-\kappa L)\right],
\ee
for $\e_m=0$ ($s=+1$), where we rescaled the electrostatic free energy by the characteristic energy
\be\label{char}
\Delta\Omega^*=k_BT\frac{\ell_B\lambda^2}{2\kappa}.
\ee
We can see that the free energy given by Eq.~(\ref{cost3}) increases steadily with the polymer length up to $L\approx \kappa^{-1}$ and converges towards the saturation value $\lim_{L,d\to\infty}\Delta\Omega_{\rm pol}(0)=s\Delta\Omega^*$ beyond which the electrostatic free energy does not depend on the polymer length. The relation of Eq.~(\ref{char}) shows that for long polymers $\kappa L\gg1$, the electrostatic free energy at the membrane surface scales with ion density as $\Delta\Omega_{\rm pol}(0)\propto\rho_b^{-1/2}$. 

In order to get further analytical insight into the length dependence of the electrostatic free energy at the membrane surface, we Taylor expand Eq.~(\ref{cost3}). We find that for dilute electrolytes or short polymers $\kappa L\ll1$, the free energy increases linearly with length,
\be\label{cost4}
\Delta\Omega_{\rm pol}(0)=s\ln(2)k_BT\lambda^2\ell_B L +O\left[(\kappa L)^2\right].
\ee
At large lengths or in strong salt solutions $\kappa L\gg1$, the electrostatic free energy reaches exponentially fast the strict thermodynamic limit of Eq.~(\ref{char}),
\bea\label{cost5}
\Delta\Omega_{\rm pol}(0)&=&sk_BT\frac{\ell_B\lambda^2}{2\kappa}\left\{1+\frac{2}{\kappa L}\left(\frac{2}{\kappa L}-1\right)e^{-\kappa L}\right\}\nonumber\\
&&+O\left(e^{-2\kappa L}\right).
\eea
Moreover, defining the saturation condition of the free energy as $\Delta\Omega_{\rm pol}(z_t=0)\gtrsim0.9\Delta\Omega^*$, we find that the former saturates at $\kappa L\gtrsim2$. This yields the characteristic polymer length determining the thermodynamic limit $L^*=2/\kappa$. We plot the latter equality in the inset of Fig.~\ref{fig3}. We see that the higher the salt concentration, the smaller the thermodynamic length. Indeed, we find $L^*\approx20$ nm (equivalent to a $\approx 100$ bps dsDNA sequences) at the salt density $\rho_b=10^{-3}$ M, $L^*\approx6$ nm ($\approx 30$ bps) for $\rho_b=10^{-2}$ M, and $L^*\approx2$ nm ($\approx10$ bps) at $\rho_b=10^{-1}$ M. It is noteworthy that beyond these critical lengths where finite size effects are irrelevant, the electrostatic free energy of Eq.~(\ref{cost2}) takes for $L \to \infty$ a much simpler form
\be\label{cost7}
\Delta\Omega_{\rm pol}(z_t)=sk_BT\frac{\ell_B\lambda^2}{2\kappa}\left[e^{-2\kappa \vert z_t \vert}+2\kappa \vert z_t \vert\;\mathrm{Ei}(-2\kappa \vert z_t \vert)\right].
\ee

After having investigated the short distance behaviour of the electrostatic free energy, we now consider its large distance behaviour. By Taylor-expanding Eq.~(\ref{cost2}) in the regime $|\kappa z_t|\gg1$, we find to leading order
\bea
\label{cost8}
\Delta\Omega_{\rm pol}(z_t)&\approx& sk_BT\frac{\ell_BQ_{\rm eff}^2(L)}{4 \vert z_t \vert}e^{-2\kappa \vert z_t \vert}.
\eea 
In Eq.~(\ref{cost8}), we introduced the effective polymer charge
\be
\label{efq}
Q_{\rm eff}(L)=\lambda L\;\frac{1-e^{-\kappa L}}{\kappa L}.
\ee
Interestingly, Eq.~(\ref{cost8}) has exactly the form of the image-charge potential experienced by a point ion of valency $Q_{\rm eff}(L)$ located at the distance $-z_t$ from a dielectric interface~\cite{Buyuk2010}. 
Equations (\ref{cost8})-(\ref{efq}) indicate that in dilute salt solutions or for short sequence lengths, polymers far away from the membrane interact with the latter as point charges with valency $Q_{\rm eff}(L\ll\kappa^{-1})=L\lambda$. Thus, in this physical regime, polymer-membrane interactions are governed by the bare polymer charge. In the opposite case of long DNA sequences or strong salt, the effective charge takes the form $Q_{\rm eff}(L\gg\kappa^{-1})=\lambda/\kappa$, indicating that the intensity of the interactions is set by the net charge of the polymer dressed by the surrounding counterion cloud.

Since the salt concentration is an easily controllable parameter in translocation experiments, it is important to characterize the influence of salt on the range and the magnitude of the polymer free energy. In the inset of Fig.~\ref{fig2II} where we plot Eq.~(\ref{costIII0}) (solid red curve), we see that the lower the salt concentration, the larger the electrostatic free energy at the membrane surface. More precisely, the reduction of the ion density from $\rho_b=10^{-1}$ M to $10^{-3}$  increases the free energy by an order of magnitude from $\approx 10$ $k_BT$ to $\approx 100$ $k_BT$.  In order to consider the range of the interactions, we remove finite size effects and focus on the limit $L\to\infty$.  By Taylor expanding Eq.~(\ref{cost7}) for large distances $|\kappa z_t|\gg1$, we get the electrostatic free energy in the asymptotic form
\be
\label{cost9}
\Delta\Omega_{\rm pol}(z_t)\approx sk_BT\frac{\ell_B\lambda^2}{4\kappa^2 \vert z_t \vert}e^{-2\kappa \vert z_t \vert}=s\Delta\Omega^*\frac{e^{-2\kappa \vert z_t \vert}}{2\kappa \vert z_t \vert}.
\ee
In the second equality of Eq.~(\ref{cost9}), we have separated the surface free energy barrier of Eq.~(\ref{char}) and the Yukawa type of decay function $e^{-2\kappa \vert z_t \vert}/2\kappa \vert z_t \vert$. Numerically, we find that this function reduces the energy by an order of magnitude at the distance $2\kappa \vert z_t \vert \approx 2$, which fixes the characteristic range of the interaction as $z^*=\kappa^{-1}$. This equality yields $z^*\approx1.0$ nm for $\rho_b=10^{-1}$ M (see also Fig.~\ref{fig2}), $z^*\approx3.0$ nm at $\rho_b=10^{-2}$ M, and $z^*\approx10$ nm at $\rho_b=10^{-3}$ M. Therefore, the reduction of the salt density significantly increases the range of polymer-membrane interactions. 

Before concluding, we consider the range of polymer-membrane interactions in a pure solvent. Neglecting the screening parameter $\kappa$, taking the large pore limit $d\to\infty$, and introducing the reduced separation distance $\bz_t=
\vert z_t \vert/L$, we can carry out the integral of Eq.~(\ref{cost2}) and get
\bea
\label{cost9II}
\Delta\Omega_{\rm pol}(z_t)&=&k_BT\ell_BL\lambda^2\Delta_0\left\{\ln\left[\frac{2+2\bz_t}{1+2\bz_t}\right]\right.\\
&&\hspace{2.2cm}\left.-\bz_t\ln\left[\frac{(1+2\bz_t)^2}{4\bz_t(\bz_t+1)}\right]\right\},\nonumber
\eea
with the salt-free dielectric discontinuity parameter $\Delta_0=(\e_w-\e_m)/(\e_w+\e_m)$. We now note that at large separation distances $|z_t|\gg L$, the free energy~(\ref{cost9II}) decays algebraically as
\be\label{cost9IV}
\Delta\Omega_{\rm pol}(z_t)\approx k_BT\Delta_0\frac{\ell_B\left(\lambda L\right)^2}{4\vert z_t \vert}.
\ee
Equation~(\ref{cost9IV}) has the form of the image potential of a point charge with valency $Q_{\rm eff}=\lambda L$ located at a distance $\vert z_t \vert$ from a dielectric interface~\cite{Buyuk2010}. The form of this free energy indicates that in  pure solvents or dilute electrolytes with $\kappa L\ll1$, the range of polymer-membrane interactions is set by the Bjerrum length $\ell_B$. In other words, the charge screening is replaced by the dielectric screening. Next, we investigate the effect of the membrane thickness on the strength of these interactions.

\subsubsection{Membrane thickness $d$}
\label{dvsL}

\begin{figure}
\includegraphics[width=1.2\linewidth]{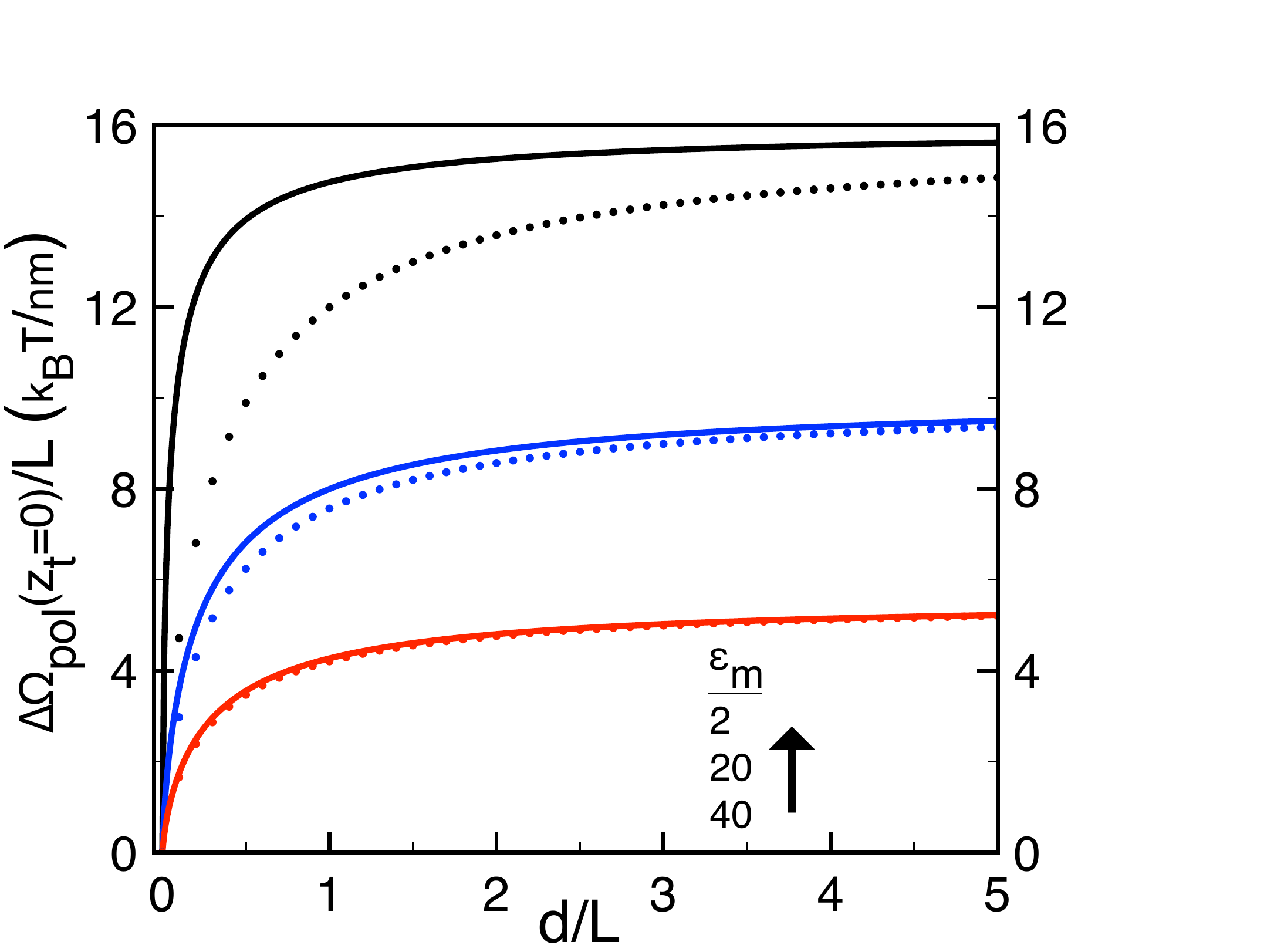}
\caption{(Color online) Electrostatic free energy density of Eq.~(\ref{cost10}) at the membrane surface versus the ratio $d/L$ in pure solvents ($\rho_b=0$) for various membrane permittivities displayed in the legend (solid curves). Dotted curves display the closed-form expression of Eq.~(\ref{cost11}). The model parameters are the same as in Fig.~\ref{fig2}.}
\label{fig5}
\end{figure}

Artificial membranes used in translocation experiments possess various thicknesses ranging from $d=6$ {\AA} for graphene-based membranes~\cite{Garaj} to $d=250$ nm for Si-based membranes~\cite{e7}. Motivated by this fact, we investigate herein the effect of the membrane thickness $d$ on the electrostatic polymer free energy. We first consider the salt-free limit $\rho_b\to0$ of pure solvents. To this end, we set in Eq.~(\ref{cost1}) $z_t=0$ and $\kappa=0$. Introducing again the salt-free dielectric discontinuity parameter $\Delta_0=(\e_w-\e_m)/(\e_w+\e_m)$ and the new integration variable $q=kL$, the free energy of Eq.~(\ref{cost1}) becomes
\bea\label{cost10}
\frac{\Delta\Omega_{\rm pol}(0)}{k_BT L}&=&\frac{\Delta_0\ell_B\lambda^2}{2}\int_0^\infty\frac{\mathrm{d}q}{q^2}\frac{\left(1-e^{-2qd/L}\right)}{1-\Delta_0^2e^{-2qd/L}}\nonumber\\
&&\hspace{2.2cm}\times\left(1-e^{-q}\right)^2.
\eea
The integral term of Eq.~(\ref{cost10}) accounting for finite size effects depends solely on the ratio $d/L$. This indicates that finite size effects are governed by the competition between the pore thickness and the polymer length. We plot the electrostatic free energy per length in Eq.~(\ref{cost10}) in Fig.~\ref{fig5}.  Due to the strengthtening of the image interactions, the amplitude of the free energy at the membrane surface increases with the membrane thickness $d$ from zero to the saturation value
\be\label{as}
\lim_{d\to\infty}\Delta\Omega_{\rm pol}(0)\approx\Delta_0k_BT\ell_BL\lambda^2\ln(2).
\ee

In order to explain the non-linear shape of the free energy curves in Fig.~\ref{fig5}, one can derive an approximate closed-form expression. To this aim, we carry out the integral in Eq.~(\ref{cost10}) by neglecting the function in the denominator, which consists in considering exclusively the first dielectric images. Introducing the adimensional pore size $\bd=d/L$ to simplify the notation gives
\be
\label{cost11}
\frac{\Delta\Omega_{\rm pol}(0)}{k_BTL}\approx\frac{\Delta_0\ell_B\lambda^2}{2}\left\{\ln\left[\frac{1+2\bd}{1+\bd}\right]+\bd\ln\left[\frac{(1+2\bd)^2}{4\bd(1+\bd)}\right]\right\}.
\ee
In Fig.~\ref{fig5}, we show that this analytic formula reproduces the result of the integral relation of Eq.~(\ref{cost10}) with quantitative accuracy for moderate dielectric discontinuities and qualitatively for strong dielectric jumps. According to Eq.~(\ref{cost11}), for membranes with thickness much smaller than the polymer length $d\ll L$, the electrostatic free energy grows 
linearly with the ratio $d/L$ as
\be
\Delta\Omega_{\rm pol}(0)\approx\Delta_0k_BT\ell_BL\lambda^2\left\{1-\ln\left(\frac{4d}{L}\right)\right\}\frac{d}{L},
\ee
while for thick membranes $d\gg L$, the free energy converges towards the asymptotic value of Eq.~(\ref{as}) according to the inverse algebraic relation
\be
\Delta\Omega_{\rm pol}(0)\approx\Delta_0k_BT\ell_BL\lambda^2\left\{\ln(2)-\frac{L}{4d}\right\}.
\ee
Figure \ref{fig5} indicates that the saturation sets in between these two regimes at $d\approx L$. Thus, in pure solvents, finite pore size effects are negligible as long as the pore thickness is larger than the polymer length.

\begin{figure}
\includegraphics[width=1.15\linewidth]{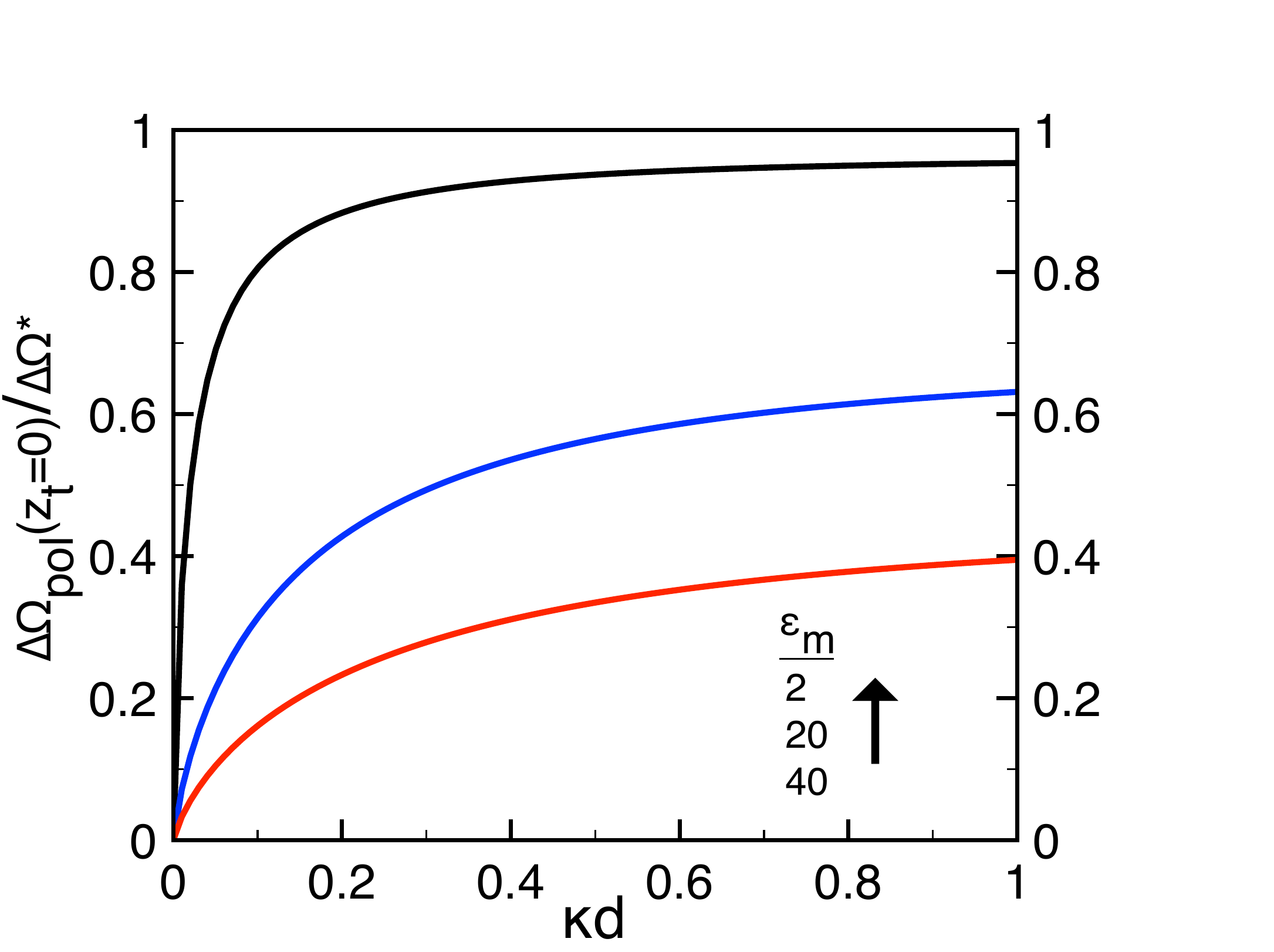}
\caption{(Color online) Electrostatic free energy of Eq.~(\ref{cost12}) at the membrane surface rescaled by the characteristic free energy $\Delta\Omega^*$ of Eq.~(\ref{char}) versus the ratio $\kappa d$ for different membrane permittivities displayed in the legend. The model parameters are the same as in Fig.~\ref{fig2}.}
\label{fig6}
\end{figure}

We investigate next the effect of the membrane thickness on membrane-polymer interactions at finite salt density. In order to simplify the analysis, we consider the thermodynamic limit $\kappa L\to\infty$ scrutinized in Section~\ref{LvsRHO} and sketched in Fig.~\ref{fig3}. Introducing the adimensional wave vector $q=k/\kappa$ and setting $z_t=0$, the electrostatic free energy of Eq.~(\ref{cost1}) rescaled with Eq.~(\ref{char}) becomes
\be\label{cost12}
\frac{\Delta\Omega_{\rm pol}(0)}{\Delta\Omega^*}=\int_0^\infty\frac{\mathrm{d}qq}{\bar{p}^3}\frac{\bar{\Delta}\left(1-e^{-2q\kappa d}\right)}{1-\bar{\Delta}^2e^{-2q\kappa d}},
\ee
where we introduced the adimensional parameters $\bar{p}=\sqrt{1+q^2}$ and 
$\bar\Delta=(\bar{p}-\gamma q)/(\bar{p}+\gamma q)$. In Fig.~\ref{fig6}, the plot of Eq.~(\ref{cost12}) shows that the increase of the adimensional thickness $\kappa d$ is accompanied by the rise of the electrostatic free energy towards the upper boundary determined by Eq.~(\ref{costIII0}). Thus, the lower the salt density, the more pronounced the finite membrane size effects. Moreover, at given salt density the stronger the dielectric contrast, the smaller the characteristic membrane thickness where the free energy saturates. 

In order to quantitatively determine the physical conditions where finite membrane size matters, we calculate with Eq.~(\ref{cost12}) the characteristic membrane size $d^*$ where the electrostatic free energy saturates. We find that at the permittivity $\e_m=2$ of carbon-based membranes, the saturation of the function $\Delta\Omega_{\rm pol}(0)/\Delta\Omega^*$ occurs at $\kappa d^*\approx0.165$. This yields $d^*\approx2$ {\AA} at the salt density $\rho_b=0.1$ M, $d^*\approx5$ {\AA} for $\rho_b=0.01$ M, and $d^*\approx1.6$ nm at $\rho_b=0.001$ M. These values indicate that in DNA translocation experiments, even the thinnest graphene-based membranes of thickness $d=6$ {\AA}~\cite{Garaj} can be considered in the thermodynamic regime $\kappa d\to\infty$ as long as the salt density is above the value $\rho_b\approx 0.01$ M. This is shown in the inset of Fig.~\ref{fig2II} where we compare the electrostatic free energy at the surface of a membrane with finite thickness $d=6$ {\AA} (dashed curve) and in the limit $d\to\infty$ (solid curve). One sees that finite size effects become indeed noticeable for $\rho_b\lesssim0.01$ M but the electrostatic energy barrier $\Delta\Omega_{\rm pol}(0)\approx 80$ $k_BT$ still remains very large in this density regime.  This shows that polymer-membrane interactions induced by dielectric images are relevant even for sub-nanometer membrane thicknesses.

\subsection{Charged membranes}
\label{charged}

Depending on the pH of the solution, membrane surfaces subject to protonation processes may possess a finite average charge distribution. Motivated by this fact we consider now the coupling between the polymer and the membrane charge. This is taken into account by the second term of Eq.~(\ref{fr2}). As we found that finite membrane size corrections are irrelevant in physiological conditions, we take the infinitely thick membrane limit $d\to\infty$. Injecting the structure factor of Eq.~(\ref{str}) into Eq.~(\ref{fr2}) together with the Fourier-transformed Green's functions~(\ref{ker3})-(\ref{ker5}), we get after some algebra the electrostatic free energy in the form
\bea
\label{cost12II}
\frac{\Delta\Omega_{\rm pol}(z_t)}{k_BT}&=&\frac{\ell_B\lambda^2}{2}\int_0^\infty\frac{\mathrm{d}kk}{p^3}\Delta\left(1-e^{-pL}\right)^2e^{- 2p \vert z_t \vert}\nonumber\\
&&-\frac{2Q_{\rm eff}(L)}{\mu\kappa}\;e^{- \kappa \vert z_t \vert}.
\eea
In Eq.~(\ref{cost12II}), we introduced the Gouy-Chapman length $\mu^{-1}=2\pi\ell_B\sigma_s$ and used the effective polyelectrolyte charge $Q_{\rm eff}(L)$ of Eq.~(\ref{efq}). 
\begin{figure}
\includegraphics[width=1.2\linewidth]{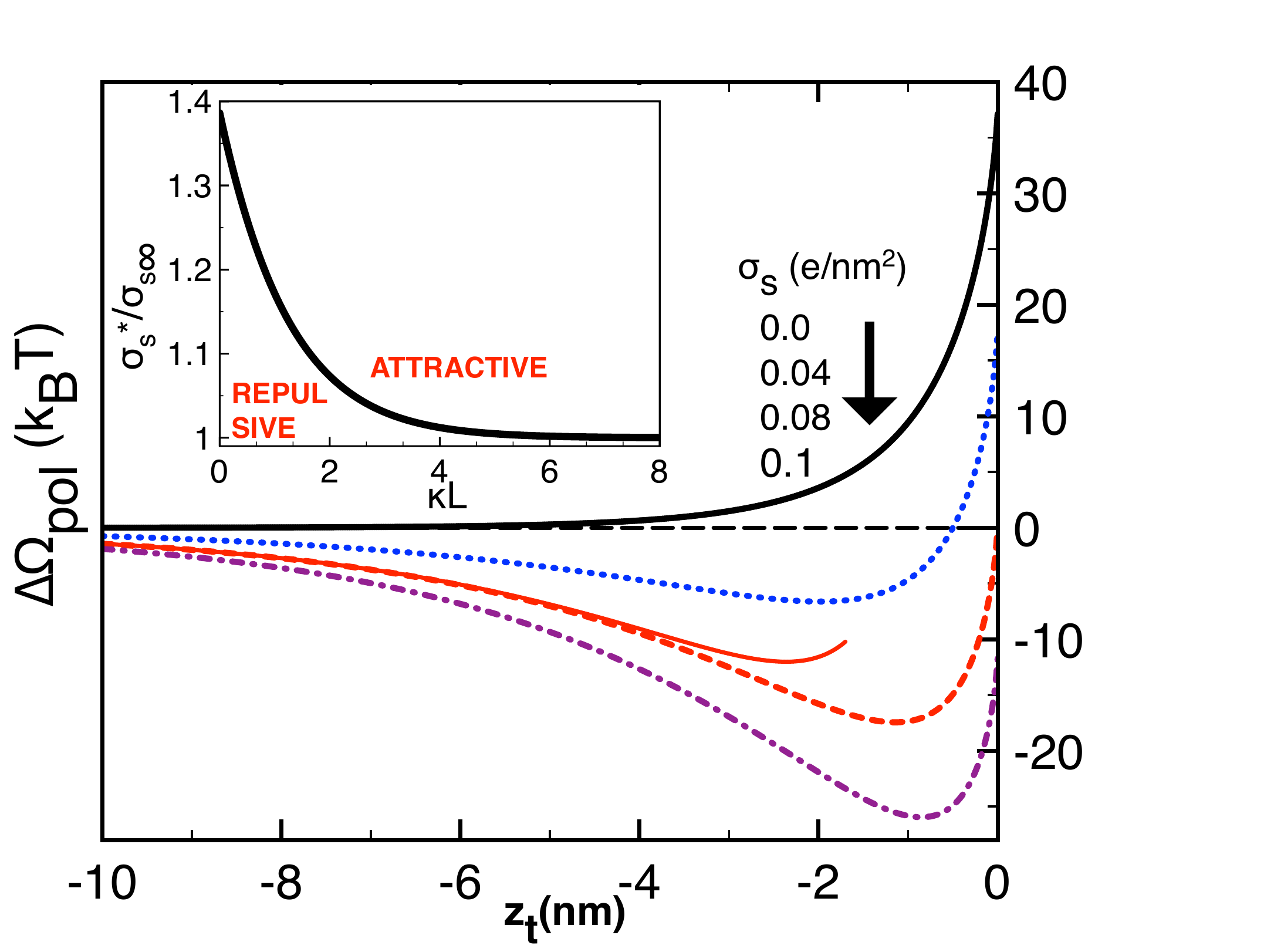}
\caption{(Color online) Main plot: Electrostatic free energy profile (Eq.~(\ref{cost13})) at membrane permittivity $\e_m=0$,  polymer length $L=1.0$ $\mu$m, salt density $\rho_b=10^{-2}$ M, and various surface charges as displayed in the legend. The solid red curve is from the asymptotic large distance law of Eq.~(\ref{cost13II}).  Inset: Characteristic surface charge (Eq.~(\ref{cr})) separating the attractive and repulsive membrane regimes rescaled by the long polymer limit of Eq.~(\ref{l}) against the adimensional polymer length $\kappa L$. The remaining model parameters are the same as in Fig.~\ref{fig2}.}
\label{fig7}
\end{figure}
\subsubsection{Membrane charge $\sigma_s$}

In order to understand the influence of the membrane charge on the electrostatic polymer free energy, we focus on the most relevant case of very low and very large permittivity membranes (see Eqs.~(\ref{con1})-(\ref{con2})). Within this restriction, the free energy of Eq.~(\ref{cost12II}) takes the form
\be
\label{cost13}
\frac{\Delta\Omega_{\rm pol}(z_t)}{k_BT}=s\frac{\ell_B\lambda^2}{2\kappa}G(z_t)-\frac{2Q_{\rm eff}(L)}{\mu\kappa}\;e^{- \kappa \vert z_t \vert},
\ee
with the function $G(z_t)$ introduced in Eq.~(\ref{g}).  We consider a positively charged membrane $\sigma_s\geq0$ of low permittivity $\e_m\ll\e_w$ and set $s=+1$. In Fig.~\ref{fig7}, we plot the electrostatic free energy profile of 
Eq.~(\ref{cost13}) at salt density $\rho_b=0.01$ M for various membrane charges up to $\sigma_s=0.1$ $e/\mbox{nm}^2$. Due to the attractive term on the r.h.s. of Eq.~(\ref{cost13}), increasing membrane charge results in lowering of the free energy which
eventually switches from positive to negative. More precisely, it acquires a negative branch associated with a minimum located at $z_t\approx-1$ nm. We see that for the largest value $\sigma_s=0.1$ $e/\mathrm{nm}^2$ which still corresponds to a weakly charged membrane, the depth of the free energy well is significantly large at about $-25$ $k_BT$. In translocation experiments, the presence of such a deep well may allow to control the approach velocity of DNA by tuning the chemical properties of the membrane surface.
\begin{figure}
\includegraphics[width=1.1\linewidth]{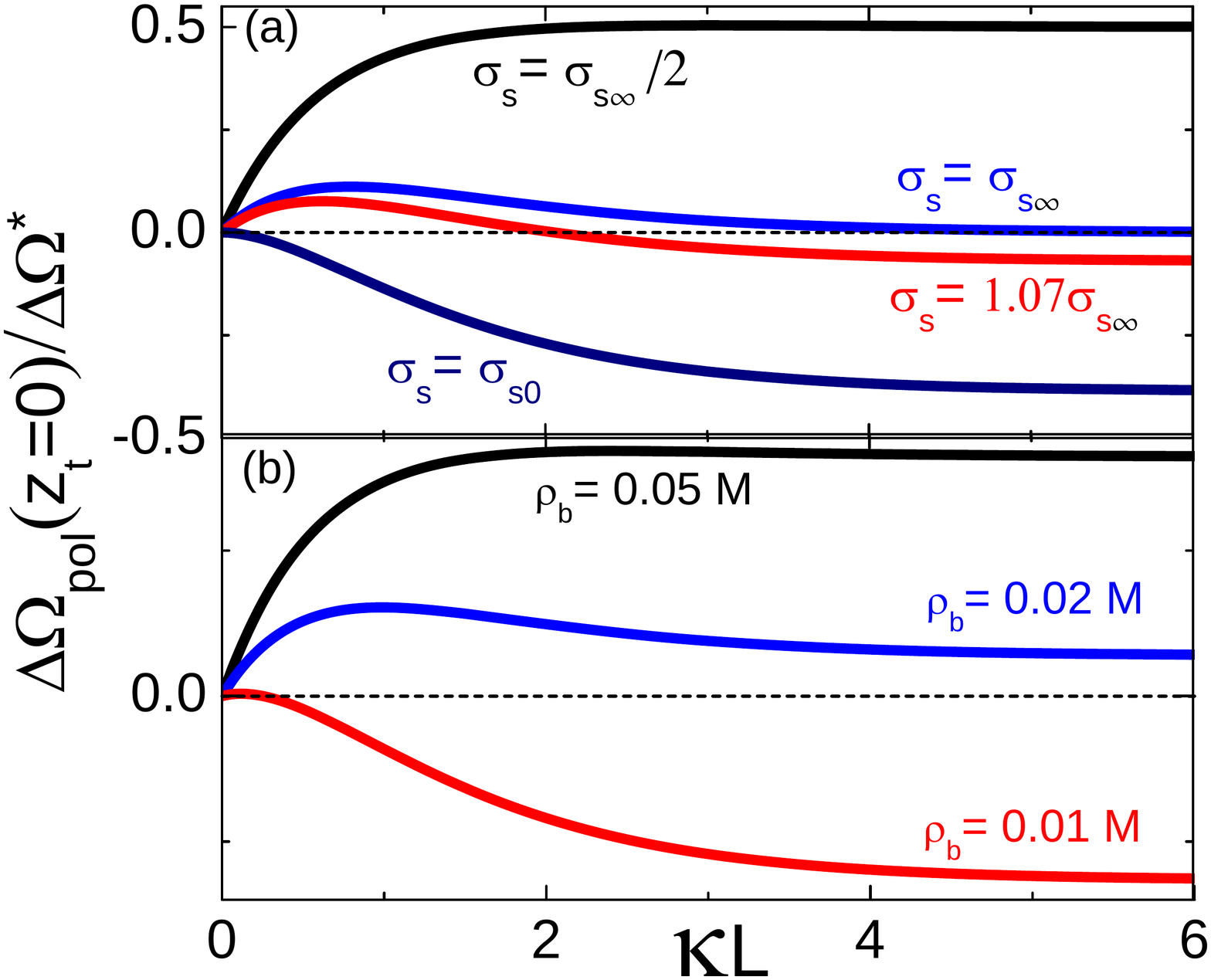}
\caption{(Color online) Electrostatic free energy of Eq.~(\ref{cost14}) at the membrane surface rescaled by the characteristic free energy $\Delta\Omega^*$ of Eq.~(\ref{char}) versus the adimensional polymer length $\kappa L$ for various (a) surface charges  and (b) salt densities at the permittivity $\e_m=0$. In (a), the salt density is $\rho_b=0.01$ M and in (b), the surface charge is $\sigma_s=0.1$ $e/\mathrm{nm}^2$.}
\label{fig8}
\end{figure}

Next, we focus on the large distance behaviour $\kappa z_t\gg1$ of the electrostatic free energy of Eq.~(\ref{cost13}). To leading order, the latter takes the form
\be
\label{cost13II}
\frac{\Delta\Omega_{\rm pol}(z_t)}{k_BT}=\frac{\ell_BQ^2_{\rm eff}(L)}{4|z_t|}e^{-2\kappa |z_t|}-\frac{2Q_{\rm eff}(L)}{\kappa\mu}e^{-\kappa |z_t|}.
\ee
This free energy has exactly the form of the net electrostatic potential of a point charge $Q_{\rm eff}(L)$ located at the distance $z_t$ from a charged dielectric wall~\cite{Buyuk2010}. The functional form of Eq.~(\ref{cost13II}) plotted in Fig.~\ref{fig7} (solid red curve) explains the negative sign of the polymer free energy far away from the surface: the polymer-surface charge attraction (the second term), being longer ranged than the image charge repulsion (the first term), dominates the repulsive image interactions at large separation distances. This means that in the presence of a finite surface charge, the polymer free energy will always have an attractive branch far enough from the interface.

\subsubsection{Polymer length $L$ and salt density $\rho_b$}

In order to consider the influence of the polymer length on the electrostatic free energy of the DNA close to a charged membrane, we investigate the short distance behaviour of polymer-membrane interactions. The form of the free energy at the membrane surface for the permittivity $\e_m=0$
\be
\label{cost14}
\frac{\Delta\Omega_{\rm pol}(0)}{k_BT}=\frac{\ell_B\lambda^2}{2\kappa}G(0)-\frac{2Q_{\rm eff}(L)}{\kappa\mu},
\ee
suggests that there exists a characteristic membrane charge $\sigma_s^*$ where the free energy on the surface vanishes. In Fig.~\ref{fig7}, this corresponds to the dashed red curve at $\sigma_s=0.08$ $e/\mathrm{nm}^2$. By equating Eq.~(\ref{cost14}) to zero and inverting the relation, the critical charge can be expressed as
\be\label{cr}
\sigma_s^*=\frac{\kappa\lambda}{8\pi}\frac{\left(1-e^{-\kappa L}\right)^2+2\kappa L\left[\mathrm{Ei}(-2\kappa L)-\mathrm{Ei}(-\kappa L)\right]}{1-e^{-\kappa L}}.
\ee
We plot Eq.~(\ref{cr}) in the inset of Fig.~\ref{fig7}. Increasing the reduced polymer length $\kappa L$, the critical charge drops smoothly from
\be\label{s}
\sigma_{s0}\approx2\ln(2)\frac{\kappa\lambda}{8\pi},
\ee
for $\kappa L\ll1$ to
\be\label{l}
\sigma_{s\infty}\approx\frac{\kappa\lambda}{8\pi},
\ee
for $\kappa L\gg1$. We note that in both limits the characteristic charge is independent of the polymer length.

We consider next the length dependence of the electrostatic free energy of Eq.~(\ref{cost14}). For short polymers $\kappa L\ll1$, it takes the asymptotic form
\be
\label{cost15}
\frac{\Delta\Omega_{\rm pol}(0)}{\Delta\Omega^*}\approx2\ln(2)\left[1-\frac{\sigma_s}{\sigma_{s0}}\right]\kappa L
\ee
that switches from repulsive to attractive at $\sigma_s=\sigma_{s0}$. For long polymers $\kappa L\gg1$, the free energy reads
\be
\label{cost16}
\frac{\Delta\Omega_{\rm pol}(0)}{\Delta\Omega^*}\approx1-\frac{\sigma_s}{\sigma_{s\infty}},
\ee
which turns from positive to negative at  $\sigma_s=\sigma_{s\infty}$. This is illustrated in Fig.~\ref{fig8}(a), where we plot Eq.~(\ref{cost14}) versus $\kappa L$. Reducing the membrane charge from  $\sigma_s=\sigma_{s\infty}/2$ to $\sigma_s=\sigma_{s\infty}$, the long polymer limit drops to zero while the electrostatic free energy remains repulsive ($\Delta\Omega_{\rm pol}(0)>0$) for short polymers $L \approx \kappa^{-1}$. At the larger charge value  $\sigma_s=1.07\sigma_{s\infty}$, the free energy at the membrane surface is repulsive for short polymers, but attractive ($\Delta\Omega_{\rm pol}(0)<0$) for long polymers.  Increasing the membrane charge to  $\sigma_s=\sigma_{s0}>\sigma_{s\infty}$, in agreement with Eq.~(\ref{cost16}), polymer-membrane interactions become attractive for all polymer lengths. 

In Fig.~\ref{fig8}(b), we also consider the influence of salt. The increase of the salt density switches the polymer-membrane interaction from attractive to repulsive. Indeed, inverting the limiting laws of Eqs. (\ref{s})-(\ref{l}), we find that the critical screening parameter where the free energy on the surface switches from negative to positive is $\kappa_0=8\pi\sigma_s/[2\ln(2)\lambda]$ for short polymers ($\kappa L\ll1$) and $\kappa_\infty=8\pi\sigma_s/\lambda$ for long polymers ($\kappa L\gg1$). Thus, salt weakens the relative weight of the attractive surface charge effect with respect to repulsive image-charge interactions.
\begin{figure}
\includegraphics[width=1.2\linewidth]{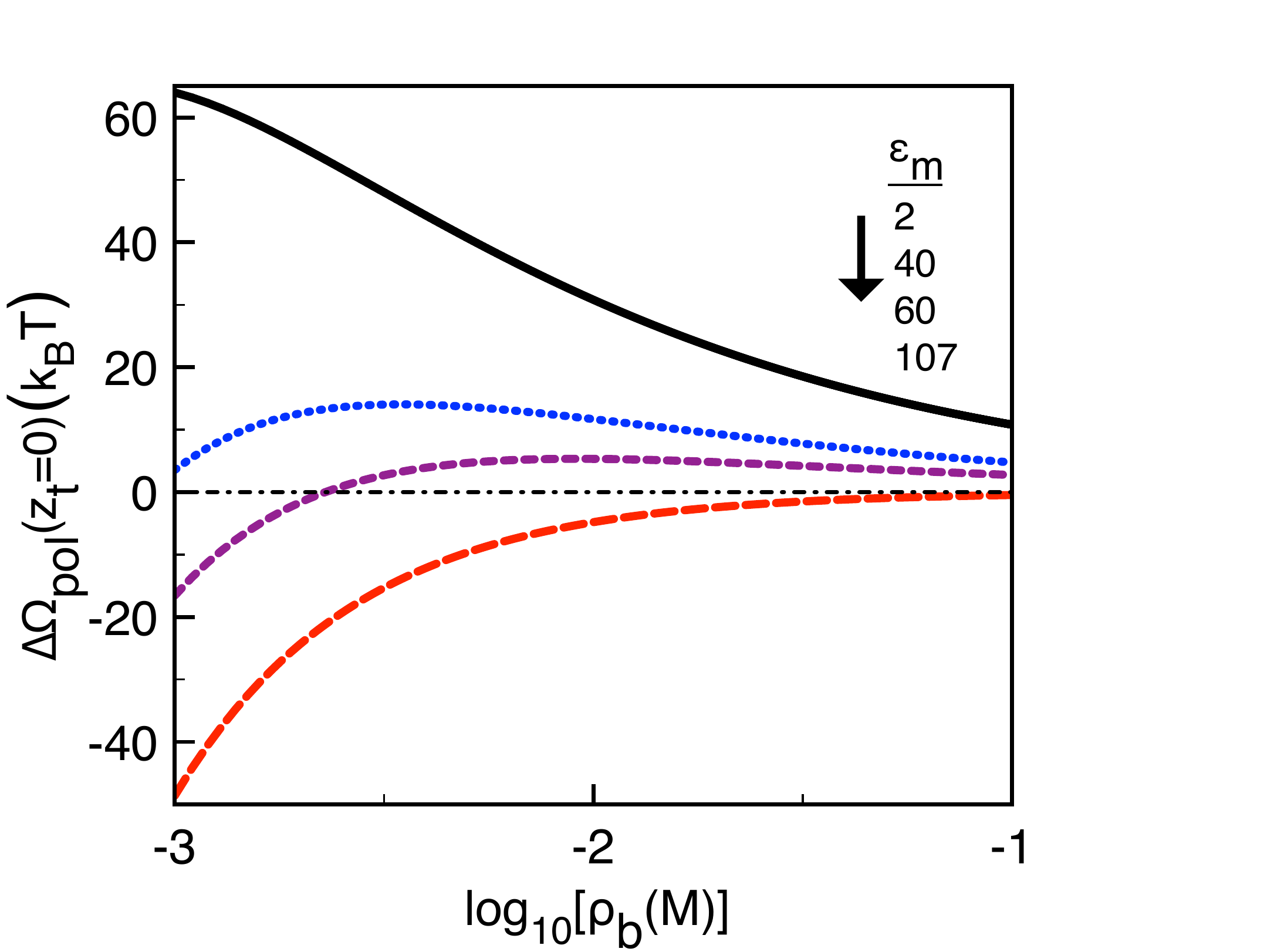}
\caption{(Color online) Electrostatic free energy at the membrane surface (Eq.~(\ref{cost17})) versus the bulk electrolyte density $\rho_b$ at the membrane charge $\sigma_s=0.01$ $e/\mathrm{nm}^2$ for various membrane permittivities given in the legend. The remaining model parameters are the same as in Fig.~\ref{fig2}.}
\label{fig9}
\end{figure}
\subsubsection{Membrane permittivity $\e_m$}

Finally, we scrutinize the influence of the membrane permittivity. Taking the limit $d\to\infty$ and $L\to\infty$, Eq.~(\ref{cost13II}) yields the electrostatic free energy at the membrane surface in an analytic form as
\be\label{cost17}
\frac{\Delta\Omega_{\rm pol}(0)}{k_BT}=\frac{\ell_B\lambda^2}{2\kappa}F(\gamma)-\frac{2\lambda}{\kappa^2\mu}
\ee
with the parameter $\gamma=\e_m/\e_w$ and the function $F(\gamma)$ given by Eqs.~(\ref{costIII})-(\ref{costV}). In Fig.~\ref{fig9}, we plot the free energy of Eq.~(\ref{cost17}) against the bulk ion density for various membrane permittivities. For a typical permittivity value $\e_m=2$ of carbon-based membranes, where polymer-membrane interactions are governed by repulsive image forces (the first term of Eq.~(\ref{cost17})), the positive free energy drops monotonically with ion density. At the intermediate value $\e_m=40$ where image forces weaken, the electrostatic free energy remains positive but exhibits a peak at the density $\rho_b\approx3\times10^{-3}$ M, below which the energetic cost decreases. This corresponds to the physical regime where the membrane charge attraction becomes relevant. By taking the derivative of Eq.~(\ref{cost17}) with respect to the screening parameter, we find that the peak is located at the bulk concentration
\be
\label{cr1}
\rho_{b,c}=\frac{32\pi\sigma_s^2}{\ell_B\lambda^2F^2(\gamma)}.
\ee
We note that for $\e_m<107$ (i.e. $F(\gamma)>0$) and $\sigma_s>0$, this density associated with the maximum energetic barrier increases both with the membrane charge ($\sigma_s\uparrow$ $\rho_{b,c}\uparrow$) and the membrane permittivity ($\e_m\uparrow$ $\rho_{b,c}\uparrow$). Then, at the permittivity value $\e_m=60$, the polymer free energy is positive at biological salt concentrations but negative for dilute electrolytes. Finally, at the value $\e_m=107$ (and for larger permittivities), because image interactions switch from repulsive to attractive, i.e. $F(\gamma)\leq0$ (see also Fig.~\ref{fig2II}), the membrane becomes purely attractive at all salt densities. 
\begin{figure}
\includegraphics[width=1.2\linewidth]{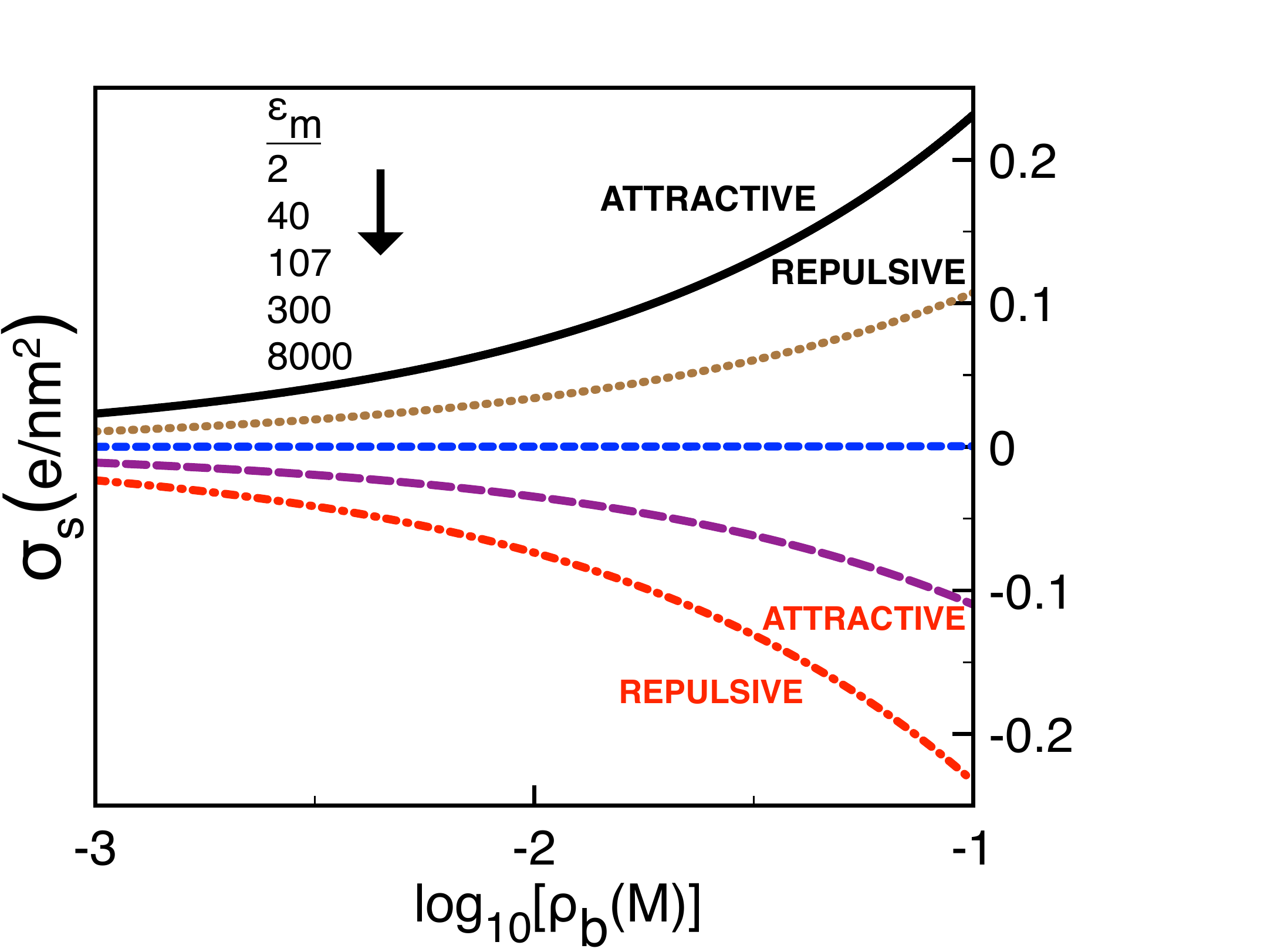}
\caption{(Color online) Phase diagram: the membrane charge (Eq.~(\ref{cr2})) versus salt density $\rho_b$ for various membrane permittivities. The characteristic curves split the regions associated with repulsive membranes (above the curves) and attractive membranes (below the curves). The remaining model parameters are the same as in Fig.~\ref{fig2}.}
\label{fig10}
\end{figure}

For translocation experiments carried out with different membrane types, it is interesting to characterize the physical regime where the free energy barrier at the membrane surface vanishes. Setting Eq.~(\ref{cost17}) to zero, we find that this occurs at the characteristic membrane charge
\be\label{cr2}
\sigma_s^*=\frac{\kappa\lambda}{8\pi}F(\gamma).
\ee
We note that Eq.~(\ref{cr2}) generalizes the limiting law of Eq.~(\ref{l}) to any permittivity value $\e_m$. Based on Eq.~(\ref{cr2}), we show in Fig.~\ref{fig10} the phase diagram characterizing the parameter regimes with attractive membranes (area above each curve) and repulsive membranes (area below each curve). In this figure, the switching of the membrane charge to negative from up to bottom stems from the fact that the attractive image forces for $\e_m>107$ have to be compensated by the repulsion between the negative membrane charge and the negative polymer charge in order for the net surface free energy to cancel out. 

The phase diagram in Fig.~\ref{fig10} indicates that at constant membrane permittivity, the larger the electrolyte density, the larger the characteristic membrane charge where the electrostatic free energy on the surface vanishes. Indeed, we have shown above that the attractive force induced by the surface charge is more susceptible to salt screening than image forces (see e.g. Fig.~\ref{fig8}(b)). Thus, a stronger salt density has to be compensated by a stronger membrane charge to cancel out the net free at the membrane surface. Furthermore, at constant salt concentration, the larger the dielectric discontinuity, the stronger the surface charge. In translocation experiments, the complex picture of this phase diagram can be at least qualitatively checked by observing the approach of a DNA molecule towards membranes with different chemical surface properties. 

\section{Summary and Conclusions}

In this work, we have developed an analytical theory accounting for electrostatic membrane-polymer interactions during the approach phase of DNA translocation events. The corresponding DH theory is beyond MF as it includes correlation effects such as image-charge interactions resulting from the dielectric mismatch between the membrane and the surrounding solvent. Within this theory, we have characterized the complex interplay between the polyelectrolyte length, the salt density, the membrane dielectric permittivity, and the membrane charge and size. 

In the first part, we considered neutral membranes. We found that in the case of thick membranes, whose permittivity strongly differs from the solvent permittivity, the approach of a long DNA molecule to the membrane costs the electrostatic free energy of magnitude $|\Delta\Omega_{\rm pol}(0)|=k_BT\ell_B\lambda^2/(2\kappa)$, where $\lambda$ is the linear DNA charge density. For neutral carbon-based membranes with low dielectric permittivity ($\e_m\approx 2$), this corresponds to a high energy barrier between $10 $ $k_BT$ to $100$ $k_BT$ depending on the salt concentration. Interestingly, the theory predicts that in the opposite case of engineered membranes with high permittivity $\e_m\gg\e_w$~\cite{Dimiev,Dang}, the membrane surface becomes an attraction point. More precisely, within the physiological salt density regime, the approach of the polymer to the membrane reduces its free energy by $10-100$ $k_BT$. We also found that in pure solvents, the electrostatic free energy becomes independent of the polymer length if the latter exceeds the membrane thickness, i.e. $L\geq d$. In electrolytes, finite size effects related to the polyelectrolyte length die out if the polymer length is larger than the DH screening length, that is $L\geq\kappa$. Most importantly, we showed that for the thinnest graphene-based membranes of thickness $d\approx 6$ {\AA}~\cite{Garaj},  the free energy barrier encountered by the DNA is close to $100$ $k_BT$. This indicates that surface polarization effects studied herein are crucial even for subnanometer membrane sizes.

In the second part, we took into account the finite charge distribution on the membrane surface. We found that even for weakly charged low permittivity membranes,  the electrostatic free energy  acquires an attractive branch far enough from the interface and turns to repulsive very close to the membrane. Because the membrane charge attraction is more sensitive to salt screening than repulsive image forces, the increase of the salt concentration makes the membrane less attractive. Furthermore, due to the competition between membrane charge and image charge effects, the sign of the polymer free energy may depend on the polymer length. We showed that for specific values of the membrane charge and ion density, the membrane will repel short polymers ($L\ll\kappa$) but attract long polymers ($L\gg\kappa$). We showed that the same competition may cancel the net electrostatic free energy on the membrane surface, which we characterized in the phase diagram of Fig.~\ref{fig10} in terms of the salt density, and the membrane charge and permittivity. This phase diagram and our general conclusions can be tested in translocation experiments.

Finally, we would like to point out limitations in the present modeling. First, our quadratic DH-level theory neglects non-linear electrostatic interactions. Hence, at low electrolyte concentrations, our free energy curves may overestimate the actual free energy barrier values. Furthermore, the rigid polymer model neglects the entropic fluctuations of the DNA molecule. These fluctuations can be considered in a future work by coupling the Coulomb liquid model with the beyond-MF formulation of the Flory theory~\cite{dun}. We also note that in the present work, we focused exclusively on the approach phase of translocation events. We would like to extend our theory to the translocation phase, consider dynamical issues, and possibly include the hydrodynamic transport in an upcoming work. We emphasize that despite the limitations of the theory, our main conclusions can be tested in translocation experiments and the theory can hopefully present itself as a starting point for more sophisticated models. The mapping between the membrane dielectric properties and the polymer free energy that we identified in this work may also allow to improve our control over DNA-membrane interactions via the chemical engineering of membrane materials.

{\acknowledgements}
This work has been supported in part by Aalto University's Energy Efficiency project EXPECTS. T.A-N. has also been supported by the Academy of Finland through its COMP Center of Excellence grant no. 251748.

\bigskip
\appendix

\section{Debye-H\"{u}ckel Free Energy} 
\label{ap1}

We present here the DH expansion of the grand potential of the electrolyte. The theory is formulated for general charge distributions in Ref.~\cite{netzvdw} and thus we will present only the general lines of the derivation. The grand canonical partition function of the charged liquid is given by the functional integral~\cite{netzvdw}
\be\label{zg}
Z_G=\int \mathcal{D}\phi\;e^{-H[\phi]},
\ee
with the Hamiltonian functional
\be\label{H}
H[\phi]=\int \mathrm{d}\br\left[\frac{\epsilon(\br)}{2\beta e^2}  [\nabla\phi(\br)]^2-i\sigma(\br)\phi(\br)-
\sum_i\lambda_i \;e^{i q_i \phi(\br)}\right],
\ee
where $\br$ stands for the position vector, $\beta=1/(k_BT)$ is the inverse temperature, $e$ the electron charge, and $\epsilon(\br)$ the dielectric permittivity function. Moreover, the function $\sigma(\br)$ accounts for immobile charge distributions in the system. The summation in the third term of Eq.~(\ref{H}) runs over the ionic species of the electrolyte, each with fugacity $\lambda_i$ and valency $q_i$. Finally, within the same field-theoretic representation, local ion densities are given by
\be\label{den1}
\rho_i(\br)=\lambda_i\lan e^{iq_i\phi(\br)}\ran_\phi,
\ee
where the bracket $\lan\cdot\ran_\phi$ denotes the average over fluctuating potential configurations taken with respect to the functional~(\ref{H}).

The DH approximation consists in Taylor expanding the functional~(\ref{H}) at the quadratic order in the fluctuating potential $\phi(\br)$. One gets the DH functional in the form
\bea\label{H0}
H_0[\phi]&=&\int \mathrm{d}\br\left[\frac{\epsilon(\br)}{2\beta e^2}  [\nabla\phi(\br)]^2-i\sigma(\br)\phi(\br)\right]-V\sum_i\lambda_i\nonumber\\
&&+\sum_i\lambda_i\int\mathrm{d}\br\left[-iq_i\phi(\br)+\frac{q_i^2}{2}\phi^2(\br)\right].
\eea
Evaluating in the bulk region the ion density~(\ref{den1}) within the same DH approximation gives
\be\label{den2}
\rho_{i,b}=\lambda_i\left[1+iq_i\lan\phi(\br)\ran_\phi-\frac{q_i^2}{2}\lan\phi^2(\br)\ran_\phi\right]_b,
\ee
where the subscript $b$ means that the field theoretic averages should be evaluated in bulk, i.e. far from any charged macromolecules breaking the spherical symmetry of the electrolyte. Noting that the average electric field should be zero in bulk, i.e. $\lan\phi(\br)\ran_{\phi,b}=0$, and inverting Eq.~(\ref{den2}) in the DH approximation gives
\be\label{den3}
\lambda_i=\rho_{i,b}\left(1+\frac{q_i^2}{2}v_{\rm DH,{\it b}}\right),
\ee
where we defined the bulk limit of the DH Green's function $v_{\rm DH,{\it b}}=\lan\phi^2(\br)\ran$ for $\br$ in the bulk region. By inserting into Eq.~(\ref{H0}) the expression for fugacity~(\ref{den3}) together with the electroneutrality condition $\sum_i\rho_{ib}q_i=0$, neglecting the terms beyond the one-loop level, and restricting ourselves to the case of a symmetric electrolyte composed of two ionic species with $\rho_{+b}=\rho_{-b}=\rho_b$ and $q_+=-q_-=q$, the Hamiltonian functional becomes
\be\label{H02}
H_0[\phi]=\int\frac{\mathrm{d}\br\mathrm{d}\br'}{2}\phi(\br)v^{-1}_{\rm DH}(\br,\br')\phi(\br')-i\int\mathrm{d}\br\sigma(\br)\phi(\br),
\ee
where we defined the DH kernel as
\be\label{ker}
v^{-1}_{\rm DH}(\br,\br')=\left[-\frac{1}{\beta e^2}\nabla\cdot\e(\br)\nabla+2\rho_b q^2\right]\delta(\br-\br').
\ee
We note that deriving Eq.~(\ref{ker}), we dropped the constant term $V\sum_i\lambda_i$ in Eq.~(\ref{H0}) and the term linear in the potential $\phi(\br)$ disappeared due to the electroneutrality condition. Computing the DH-level partition function with Eqs.~(\ref{zg}) and~(\ref{H02}) gives
\be\label{z0}
Z_0=\mathrm{det}^{1/2}\left(v_{\rm DH}\right)\;\exp\left[-\int\frac{\mathrm{d}\br\mathrm{d}\br'}{2}\sigma(\br) v_{\rm DH}(\br,\br')\sigma(\br')\right].
\ee
From the definition of the grand potential $\Omega_{\rm DH}=-k_BT\ln Z_0$, we finally get the latter as the superposition of the ionic and polymer free energies $\Omega_{\rm DH}=\Omega_{\rm ion}+\Omega_{\rm pol}$, each contribution respectively given by $\Omega_{\rm ion}=-k_BT\ln\mathrm{det}^{1/2}\left(v_{\rm DH}\right)$ and
\be\label{om1}
\Omega_{\rm pol}=k_BT\int\frac{\mathrm{d}\br\mathrm{d}\br'}{2}\sigma(\br)v_{\rm DH}(\br,\br')\sigma(\br').
\ee

\section{Electrostatic Green's Function in Slit Geometry}
\label{ap2}

In this Appendix, we explain the general lines of the inversion of the DH kernel equation~(\ref{ker}) in the planar membrane geometry depicted in Fig.~\ref{fig1}(a). Due to the plane geometry where the Green's function satisfies translational symmetry along the $x$ and the $y$ axes, i.e. $v_{\rm DH}(\br,\br')=v_{\rm DH}(z,z',\br_\pa-\br'_\pa)$, we can Fourier-expand the potential as
\be\label{ker2}
v_{\rm DH}(\br,\br')=\int\frac{\mathrm{d}^2\bk}{4\pi^2}\;e^{i\bk\cdot\left(\br_\pa-\br_\pa'\right)}\tv_{\rm DH}(z,z').
\ee
In Eq.~(\ref{ker2}), the dependence of the Fourier expanded potential on the wave vector $\bk$ is implicit. Moreover, the dielectric permittivity profile reads
\be\label{diel}
\e(z)=\e_w\theta(-z)+\e_m\theta(z)\theta(d-z)+\e_w\theta(z-d),
\ee
where $\theta(z)$ is the Heaviside step function. By inserting the expansion~(\ref{ker2}) into the kernel equation~(\ref{ker}), the latter simplifies as
\be
\left[\partial_z\e(z)\partial_z-p^2\right]\tv_{\rm DH}(z,z')=-\frac{e^2}{k_BT}\delta(z-z'),
\ee
with $p=\sqrt{k^2+\kappa^2}$ and the DH screening parameter $\kappa=8\pi q^2\ell_B\rho_b$. For the source charge located on the right side of the membrane $z'>d$, the piecewise homogeneous solution is
\bea\label{hom1}
\tv_{\rm DH}(z,z')&=&C_1 e^{pz}\theta(-z)+\left[C_2e^{kz}+C_3e^{-kz}\right]\theta(z)\theta(d-z)\nonumber\\
&&+\left[C_4e^{pz}+C_5e^{-pz}\right]\theta(z-d)\theta(z'-z)\nonumber\\
&&+C_6e^{-pz}\theta(z-z').
\eea
For the source located in the left half-space $z'<0$, the solution is given by
\bea\label{hom2}
\tv_{\rm DH}(z,z')&=&C_1 e^{pz}\theta(z'-z)\nonumber\\
&&+\left[C_2e^{pz}+C_3e^{-pz}\right]\theta(-z)\theta(z-z')\nonumber\\
&&+\left[C_4e^{kz}+C_5e^{-kz}\right]\theta(z)\theta(d-z)\nonumber\\
&&+C_6e^{-pz}\theta(z-d).
\eea
The integration constants $C_i$ with $1\leq i\leq6$ are to be determined by applying in each case the continuity of the potential $\tv_{\rm DH}(z,z')$ and the displacement function $\e(z)\partial_z\tv_{\rm DH}(z,z')$ at the boundaries $z=0$, $z=d$, and at $z=z'$.  
After somewhat tedious algebra we get
\bea
\label{ker3}
\tv_{\rm DH}(z\leq0,z'\leq0)&=&\tv_b(z-z')\\
&&+\frac{2\pi\ell_B}{p}\frac{\Delta\left(1-e^{-2kd}\right)}{1-\Delta^2e^{-2kd}}e^{p(z+z')}\nonumber,\\
\label{ker4}
\tv_{\rm DH}(z\geq d,z'\geq d)&=&\tv_b(z-z')\\
&&+\frac{2\pi\ell_B}{p}\frac{\Delta\left(1-e^{-2kd}\right)}{1-\Delta^2e^{-2kd}}e^{p(2d-z-z')}\nonumber,
\eea
and
\bea
\label{ker5}
\tv_{\rm DH}(z,z')&=&\tv_b(z-z')\\
&&+\frac{2\pi\ell_B}{p}\frac{(1-\Delta^2)e^{(p-k)d}+\Delta^2e^{-2kd}-1}{1-\Delta^2e^{-2kd}}e^{-p|z-z'|}\nonumber,
\eea
for $z'\leq0$ and $z\geq d$, or $z'\geq d$ and $z\leq0$. In Eqs.~(\ref{ker3})-(\ref{ker5}), the dielectric discontinuity function is defined as
\be
\Delta=\frac{\e_wp-\e_mk}{\e_wp+\e_mk}.
\ee
We finally note that in Eqs.~(\ref{ker3})-(\ref{ker5}),  we introduced the bulk part of the Fourier transformed DH potential
\be\label{ker6}
\tv_b(z-z')=\frac{2\pi\ell_B}{p}e^{-p|z-z'|}.
\ee

\end{document}